\documentclass[twocolumn]{aastex701}

\usepackage{color}
\usepackage{url}
\usepackage{hyperref}
\usepackage{xspace}
\usepackage{float}
\usepackage{soul}
\usepackage{amssymb}
\usepackage{bm}
\usepackage{orcidlink}
\usepackage{multirow}

\newcommand {\lum}{erg\,s$^{-1}$}

\newcommand {\beppo}{{BeppoSAX}\xspace}
\newcommand {\xmm}{{XMM--Newton}\xspace}
\newcommand {\nustar}{\textit{NuSTAR}\xspace}
\newcommand {\ixpe}{{IXPE}\xspace}
\newcommand {\nicer}{{NICER}\xspace}
\newcommand {\source}{{4U~1735$-$44}\xspace}

\begin{document}

\title{X-Ray Polarization from the Atoll \source Suggests a Low Inclination}

\shorttitle{Polarized Emission from 4U~1735$-$44}
\shortauthors{D\'iaz~Teodori et al.}

\author[0009-0002-1852-7671]{Mar\'ia~Alejandra~D\'iaz~Teodori}
\affiliation{Department of Physics and Astronomy, FI-20014 University of Turku,  Finland}
\affiliation{Nordic Optical Telescope, Rambla José Ana Fernández, Pérez 7, E-38711 Breña Baja, Spain}
\email{alejandra.m.diaz@utu.fi}
\correspondingauthor{Mar\'ia~Alejandra~D\'iaz~Teodori}

\author[0009-0009-3183-9742]{Anna Bobrikova}
\affiliation{Department of Physics and Astronomy, FI-20014 University of Turku,  Finland}
\email{anna.a.bobrikova@utu.fi}

\author[0000-0002-0642-1135]{Andrea~Gnarini}
\affiliation{NASA Marshall Space Flight Center, Huntsville, AL 35812, USA}
\affiliation{Dipartimento di Matematica e Fisica, Universit\`a degli Studi Roma Tre, via della Vasca Navale 84, 00146 Roma, Italy}
\email{andrea.gnarini@uniroma3.it}

\author[0000-0001-9442-7897]{Francesco Ursini}
\affiliation{Dipartimento di Matematica e Fisica, Universit\`a degli Studi Roma Tre, via della Vasca Navale 84, 00146 Roma, Italy}
\email{francesco.ursini2@uniroma3.it}

\author[0000-0001-9167-2790]{Sofia V. Forsblom}
\affiliation{Department of Physics and Astronomy, FI-20014 University of Turku,  Finland}
\email{sofia.v.forsblom@utu.fi}

\author[0000-0002-0983-0049]{Juri Poutanen}
\affiliation{Department of Physics and Astronomy, FI-20014 University of Turku,  Finland}
\email{juri.poutanen@utu.fi}

\author[0000-0003-2609-8838]{Alexander Salganik}
\affiliation{Department of Physics and Astronomy, FI-20014 University of Turku,  Finland}
\email{alsalganik@gmail.com}

\author[0000-0002-4622-4240]{Stefano Bianchi}
\affiliation{Dipartimento di Matematica e Fisica, Universit\`a degli Studi Roma Tre, via della Vasca Navale 84, 00146 Roma, Italy}
\email{stefano.bianchi@uniroma3.it}

\author[0000-0002-6384-3027]{Fiamma Capitanio}
\affiliation{INAF Istituto di Astrofisica e Planetologia Spaziali, Via del Fosso del Cavaliere 100, 00133 Roma, Italy}
\email{fiamma.capitanio@inaf.it}

\author[0000-0002-5817-3129]{Massimo Cocchi}
\affiliation{INAF Osservatorio Astronomico di Cagliari, via della Scienza 5, I-09047, Selargius (CA), Italy}
\email{massimo.cocchi@inaf.it}

\author[0000-0003-1533-0283]{Sergio Fabiani}
\affiliation{INAF Istituto di Astrofisica e Planetologia Spaziali, Via del Fosso del Cavaliere 100, 00133 Roma, Italy}
\email{sergio.fabiani@inaf.it}

\author[0000-0003-2212-367X]{Ruben Farinelli}
\affiliation{INAF Osservatorio di Astrofisica e Scienza dello Spazio, Via P. Gobetti 101, 40129 Bologna, Italy}
\email{ruben.farinelli@inaf.it}

\author[0000-0002-3638-0637]{Philip Kaaret}
\affiliation{NASA Marshall Space Flight Center, Huntsville, AL 35812, USA}
\email{philip.kaaret@nasa.gov} 

\author[0000-0002-3010-8333]{Jari~J.~E.~Kajava}
\affiliation{Serco for the European Space Agency (ESA), European Space Astronomy Centre, Camino Bajo del Castillo s/n, E-28692 Villanueva de la Ca\~{n}ada, Madrid, Spain}
\email{jari.kajava@gmail.com} 

\author[0000-0002-2152-0916]{Giorgio Matt}
\affiliation{Dipartimento di Matematica e Fisica, Universit\`a degli Studi Roma Tre, via della Vasca Navale 84, 00146 Roma, Italy}
\email{giorgio.matt@uniroma3.it}

\author[0000-0002-0940-6563]{Mason Ng}
\affiliation{MIT Kavli Institute for Astrophysics and Space Research, Massachusetts Institute of Technology, 77 Massachusetts Avenue, Cambridge, MA 02139, USA}
\affiliation{Department of Physics, McGill University, 3600 rue University, Montr\'{e}al, QC H3A 2T8, Canada}
\affiliation{Trottier Space Institute, McGill University, 3550 rue University, Montr\'{e}al, QC H3A 2A7, Canada}
\email{masonng@mit.edu}

\author[0000-0002-2381-4184]{Swati~Ravi}
\affiliation{MIT Kavli Institute for Astrophysics and Space Research, Massachusetts Institute of Technology, 77 Massachusetts Avenue, Cambridge, MA 02139, USA}
\email{swatir@mit.edu} 

\author[0000-0001-8916-4156]{Paolo Soffitta}
\affiliation{INAF Istituto di Astrofisica e Planetologia Spaziali, Via del Fosso del Cavaliere 100, 00133 Roma, Italy}
\email{paolo.soffitta@inaf.it}

\author[0009-0007-0537-9805]{Antonella Tarana}
\affiliation{INAF Istituto di Astrofisica e Planetologia Spaziali, Via del Fosso del Cavaliere 100, 00133 Roma, Italy}
\email{antonella.tarana@inaf.it}

\author[0000-0001-5326-880X]{Silvia Zane}
\affiliation{Mullard Space Science Laboratory, University College London, Holmbury St Mary, Dorking, Surrey RH5 6NT, UK}
\email{s.zane@ucl.ac.uk}




\begin{abstract}
X-ray polarimetry is a new tool capable of probing the geometry of accretion onto weakly magnetized neutron stars. Here we present the first X-ray spectro-polarimetric results from coordinated observations of the atoll source \source, conducted with the Imaging X-ray Polarimetry Explorer (\ixpe), \nicer, and \nustar. 
Over the 2--8 keV energy range, we obtained a marginal detection of polarization with the polarization degree of $1.4\%\pm0.7\%$ and polarization angle of $-29\degr\pm14\degr$, corresponding to a $3\sigma$ upper limit on the polarization degree of 3.5\%. The best-fit model to describe the spectrum comprises a thermal component associated with the accretion disk, a Comptonized blackbody component, and a relativistic reflection component. From the reflection model, we infer a disk inclination of $\sim$40\degr. The spectroscopic and polarimetric properties of \source are consistent with those observed in other atoll sources studied by IXPE, with its low polarization likely due to its low inclination.
\end{abstract}



\section{Introduction} 
\label{sec:intro}

Accreting weakly magnetized neutron stars (WMNSs) in low-mass X-ray binaries (LMXBs) are among the brightest X-ray sources in the sky and exhibit complex behavior that is not yet fully understood. They accrete mass via Roche-lobe overflow from a low-mass (typically $\leq 1 \, M_{\odot}$) companion, forming an accretion disk \citep{Bahramian2024}. 

The spectrum of WMNSs is usually modeled by the sum of two components: a soft thermal component ($<1$ keV) associated with the accretion disk \citep{ShakuraSunyaev73}, and a hard component resulting from Comptonization in a relatively cool plasma in the boundary layer (BL) between the disk and the NS surface
\citep{1988AdSpR...8b.135S,2001ApJ...547..355P} or the spreading layer (SL) where the accreted material spreads to higher latitudes \citep{IS1999,Suleimanov2006}. 
In addition, X-ray illumination of the accretion disk can produce a reflection component with both continuum and line features, most notably an iron emission line at 6--7 keV.  
Furthermore, WMNSs can also display Type-I X-ray bursts which are caused by thermonuclear explosions of the matter accreted from a companion onto the NS surface \citep[see, e.g.,][]{1993SSRv...62..223L}.

Beyond these common characteristics, WMNSs have historically been classified into two main categories: Z- and atoll sources \citep[see, e.g.,][]{1989A&A...225...79H}. This classification is based on the patterns these sources display in the color-color diagram (CCD) or the hardness-intensity diagram (HID), as well as their correlated spectral and timing properties. Atoll sources are characterized by a lower luminosity ($10^{36}$--$10^{37}$~\lum) compared to Z-sources ($10^{38}$~\lum, near the Eddington luminosity).

Despite all the advances that were reached in studying the WMNS with spectroscopy and timing, many fundamental questions remain open. For instance, both the classification into Z and atoll sources and the state transitions of the sources lack physical explanations. The geometry of the region between the NS surface and the inner parts of the accretion disk also remains a mystery. In this context, polarimetry is a powerful tool to examine the geometry of the surroundings of compact objects. Polarimetric properties of the emission coming from different regions of the WMNS have been predicted by \citet{Dovciak2008} and \citet{Loktev22} for the disk, by \citet{1985MNRAS.217..291L}, \citet{Gnarini22}, \citet{Farinelli24}, and \citet{Bobrikova25} for the Comptonized component, and by \citet{Matt1993} and \citet{Poutanen96} for the reflected component. 

\begin{deluxetable*}{ccccc}
\tablecaption{Log of observations of \source presented in the paper. }
\label{table:obsdates}
\tablehead{Observatory & Dates & ObsID & Instrument & Duration (ks)} 
\startdata
            \multirow{3}{*}{\ixpe} & \multirow{3}{*}{2024 Aug 31} & \multirow{3}{*}{03004001} & DU1 & 38.8    \\
              & &  & DU2 & 38.9 \\
              &  &  & DU3 & 38.9  \\
            \multirow{2}{*}{\nicer} & \multirow{3}{*}{} 2024 Aug 28 & 7700030101 & XTI & 1.8  \\
             &2024 Aug 29 &   7700030102 & XTI & 	0 \tablenotemark{a} \\
            & 2024 Aug 30 &   7700030103 & XTI & 	0.4 \\
             \multirow{2}{*}{\nustar} & \multirow{2}{*}{2024 Aug 31} & \multirow{2}{*}{31001011002} & FPMA & 33.7 \\
            & &  & FPMB & 34.2 \\
\enddata
\tablenotetext{a}{The second NICER observation, 7700030102, had no usable data due to known optical light leak issues.}
\end{deluxetable*}

Since the launch of the Imaging X-ray Polarimetry Explorer \citep[\ixpe;][]{Weisskopf2022} in December 2021, X-ray polarimetry has been actively used to study geometry of various accreting compact objects. 
Only for the WMNSs, a dozen discoveries has been made. 
For instance, in the Z-source \mbox{Cyg~X-2} \citep{Farinelli23} the polarization of the Comptonized component was aligned with the jet, putting a constraint on the geometry of the boundary region. 
For the atoll sources \mbox{GX~9+9}, \mbox{4U~1820$-$303}, and \mbox{4U~1624$-$49}, a trend of a strong dependence of polarization degree (PD) on energy was found \citep[][respectively]{Ursini2023, DiMarco23, Saade24}, while for the Z-sources \mbox{XTE~J1701$-$462}, \mbox{Sco~X-1} and \mbox{GX~5$-$1}, the common trend was a significant change in PD with the transition between the hard and soft state \citep[][respectively]{Cocchi23, LaMonaca2024,Fabiani24}. Even the upper limits on the PD obtained for \mbox{GS~1826$-$238}, \mbox{Ser~X-1}, \mbox{GX~3+1}, and \mbox{GX~9+1} \citep[][respectively]{Capitanio23, SerX1, Gnarini24, Tarana_2025} were used to constrain the inclination of these objects. 

It is worth noting that \mbox{GX~9+9}, \mbox{4U~1820$-$303}, and \mbox{4U~1624$-$49} showed clear signatures of reflection, whereas \mbox{GS~1826$-$238} did not. Reflected emission has been proposed as a key factor in the high polarization observed in these atoll sources, which could explain why GS 1826$-$238 showed no significant polarization while the three others did. However, \mbox{Ser~X-1}, \mbox{GX~3+1}, and \mbox{GX~9+1} despite also exhibiting reflection features, showed no polarization detection. This has been attributed to the low inclination of these sources.

Moreover, hints for a misalignment between NS rotation axis from the orbital axis were obtained for \mbox{Cir~X-1} \citep{Rankin2024} and \mbox{GX~13+1} \citep{Bobrikova24a, Bobrikova24b}, although alternative explanation in terms of eclipse of the scattering region by the outer part of the cold disk also exists \citep{DiMarco2025}. 
Note that both \mbox{Cir~X-1} \citep{Oosterbroek_1995, Shirey_1998} and \mbox{GX~13+1} \citep{Schnerr_2003,Fridriksson_2015} are peculiar cases that display both Z and atoll features. Similar behavior has been observed in other sources, with \citet{Ng_2024} recently reporting that the behavior of  \mbox{1A~1744$-$361} is consistent with the idea that atoll and Z sources may represent different accretion regimes.

We combine spectroscopic and polarimetric analyses to probe the geometry of a bright atoll source \source. 
It has a luminosity of around 10\% of the Eddington luminosity and is located at a distance of $5.6^{+3.7}_{-2.1}$~kpc \citep{2018AJ....156...58B}. 
It has an orbital period of $4.564\pm0.005$~h \citep{1989MNRAS.239..533C}. The inclination of the system is constrained from optical observations at 27\degr--60\degr\ \citep{2006MNRAS.373.1235C}. 
Analyzing \beppo and \xmm data, \cite{2013A&A...555A..17M} obtained an inclination closer to the upper limit of nearly 60\degr, while \cite{2020ApJ...895...45L}, using simultaneous \nicer and \nustar observations, obtained two values, approximately 42\degr\ or 57\degr, concluding that the constraint on inclination is highly model-dependent. 
The spectrum of \source has been studied extensively, with a special focus on the reflected component of the emission (see \citealt{2020ApJ...895...45L}, and references therein). The source is also known to experience regular Type I X-ray bursts \citep{1988A&A...192..147V, 2013AJ....146...60L}.

The paper is structured as follows. 
We describe the observatories used and the data reduction process briefly in Section~\ref{sec:obs}. 
We present the results of the data analysis in Section~\ref{sec:analysis}. 
We discuss the results and conclude in Section~\ref{sec:discussion}.

\section{Observations} 
\label{sec:obs}

\subsection{IXPE}

The Imaging X-ray Polarimetry Explorer \citep[\ixpe;][]{Weisskopf2022} is the first observatory dedicated to X-ray polarimetry. 
The mission consists of three identical telescopes, each featuring a gas pixel detector that is sensitive to X-ray polarization in the 2--8 keV energy range \citep{2021AJ....162..208S,2021APh...13302628B}. 
These detectors are positioned at the focal point of grazing incidence mirror assembly modules. 
\ixpe observed \source on 2024 August 31 for a total exposure time of about 39~ks for each detector unit (DU). 
The corresponding ObsIDs and exposure times are given in Table~\ref{table:obsdates}.

For the model-independent polarimetric analysis, we processed the data using the \textsc{ixpeobssim} software, version 30.6.3 \citep{Baldini2022}. 
The normalized Stokes $q$ and $u$ parameters, as well as the PD and polarization angle (PA), were obtained through the \texttt{pcube} binning algorithm \citep{2015APh....68...45K}.

To carry out the spectral and spectro-polarimetric analysis, the source and background regions were selected from the image of each of the three DUs. Source spectra and light curves were extracted from a circular region with a 114\arcsec\ radius centered on the source, while the background was selected as an annular region centered on the source spanning 180\arcsec–240\arcsec. We iteratively determined the source extraction radii from 30\arcsec\ to 180\arcsec\ in 5\arcsec\ steps in order to maximize the signal-to-noise ratio (S/N) across the entire IXPE energy range. This approach is similar to that used by \citet{Piconcelli_2004}. 

Due to the high brightness of this source, the background is negligible \citep[see][]{DiMarco2023} and was not subtracted. 
The data reduction was performed applying the weighted scheme \citep{DiMarco2022}. For this analysis we used the \textsc{heasoft} package, version 6.34 \citep{heasarc}, standard \textsc{ftools} and \textsc{xspec} \citep{Arnaud96}. 
The calibration files were obtained from the CALDB version 20240125. 

\subsection{\nustar}

The Nuclear Spectroscopic Telescope Array  \citep[\nustar;][]{harrison_2013}, is the first focusing high-energy X-ray mission, operating in the 3--79~keV energy range. 
It consists of two focal plane modules (FPMA and FPMB) with a spectral resolution of 400~eV (FWHM) at 10~keV.

\nustar observed \source on 2024 August 31 for a total of about 33~ks (see Table~\ref{table:obsdates}). 
We processed the data using \nustar Data Analysis Software \textsc{nustardas} provided under \textsc{heasoft} v 6.34. The extraction region radii were determined using the same iterative method as used with \ixpe. 
The extraction radii are 140\arcsec\ for both FPMs.
The calibration files were obtained from the CALDB version 20240812.

\subsection{\nicer}
\label{sect:nicer}

The Neutron Star Interior Composition Explorer \citep[\nicer;][]{gendreau_2016} is mounted on the International Space Station and provides exceptional spectral and timing resolution despite lacking imaging capabilities. 
Its primary instrument, the X-ray Timing Instrument (XTI), operates in the soft X-ray energy range of 0.2--12~keV.

\begin{figure*}
\centering
\includegraphics[width=0.75\columnwidth]{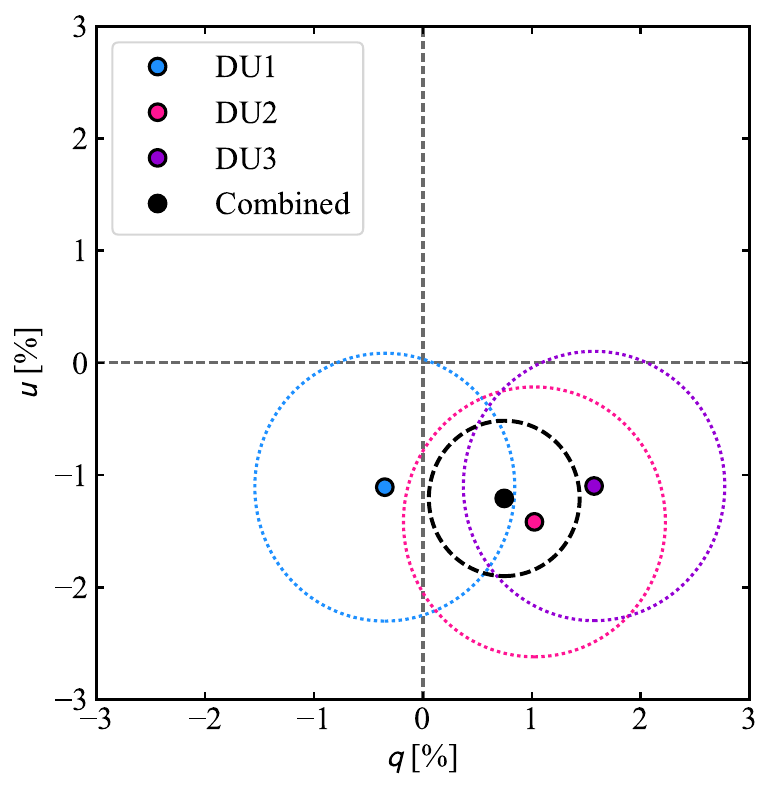}
\includegraphics[width=0.95\columnwidth]{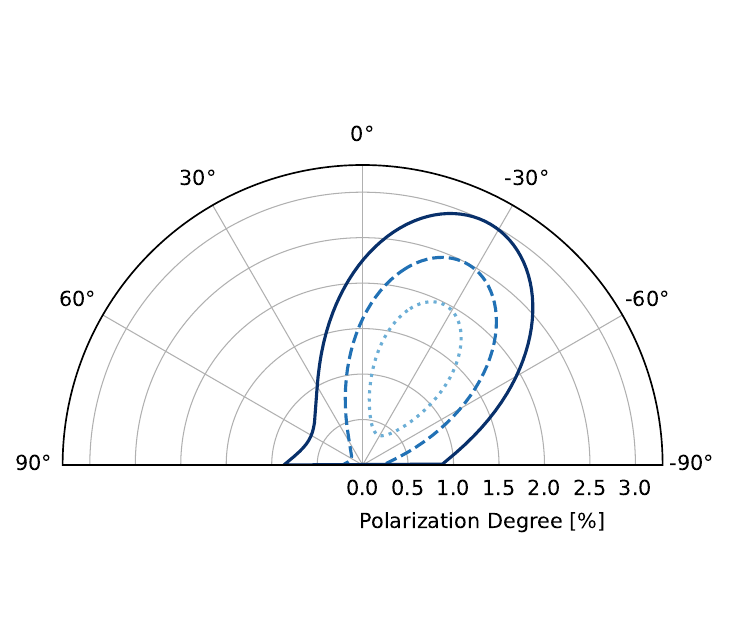}
\caption{Averaged polarization properties.
Left: Normalized Stokes $q$ and $u$ parameters in the 2--8 keV band, for the three \ixpe\ DUs and their combination. Errors are provided at $1\sigma$~CL. 
Right: Polarization contours in the 2--8 keV band at the 68\%, 90\% and 99\% confidence levels obtained with the \texttt{ixpe\_protractor} task.} 
\label{fig:averpol}
\end{figure*}

\nicer observed \source on 2024 August 28, 29, and 30 for a total of 2.2~ks (see Table~\ref{table:obsdates}), although the second NICER observation, from August 29, had no usable data due to known optical light leak issues.
We processed the data using the {NICER} Data Analysis Software $\textsc{nicerdas}$ provided under \textsc{heasoft} v 6.34. 
We used the standard {NICER}-recommended calibration and filtering tool \textsc{nicerl2} to produce the clean event files. 
Standard screening criteria were applied. 
The calibration files were obtained from the CALDB version 20240206. 

\section{Data analysis}
\label{sec:analysis}

\subsection{Model-independent polarimetric analysis} 
\label{sec:PCUBE}

\begin{figure}
\centering
\includegraphics[width=0.88\columnwidth]{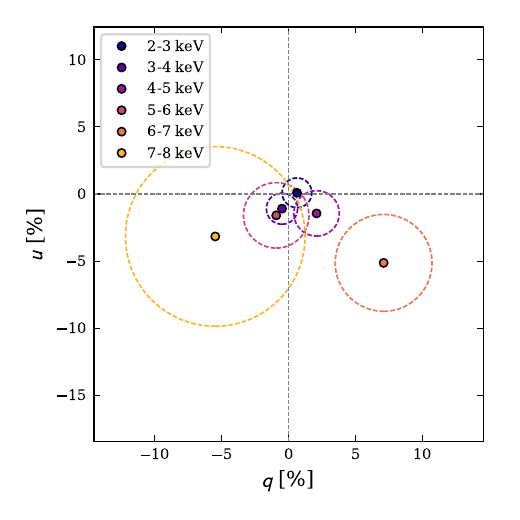}
\includegraphics[width=0.88\columnwidth]{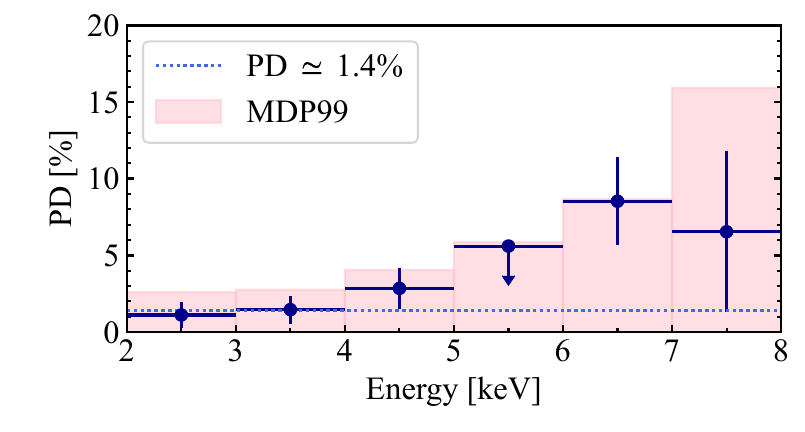}
\caption{Polarimetric properties as functions of energy. Errors are provided at $1\sigma$ CL.  
Upper panel: Normalized Stokes parameters $q$ and $u$ in six energy bands. Lower panel: PD as a function of energy. The $2\sigma$ upper limit  in the 5--6 keV energy bin is shown.}

\label{fig:polenergy}
\end{figure}

\begin{figure}
\centering
\includegraphics[width=0.98\columnwidth]{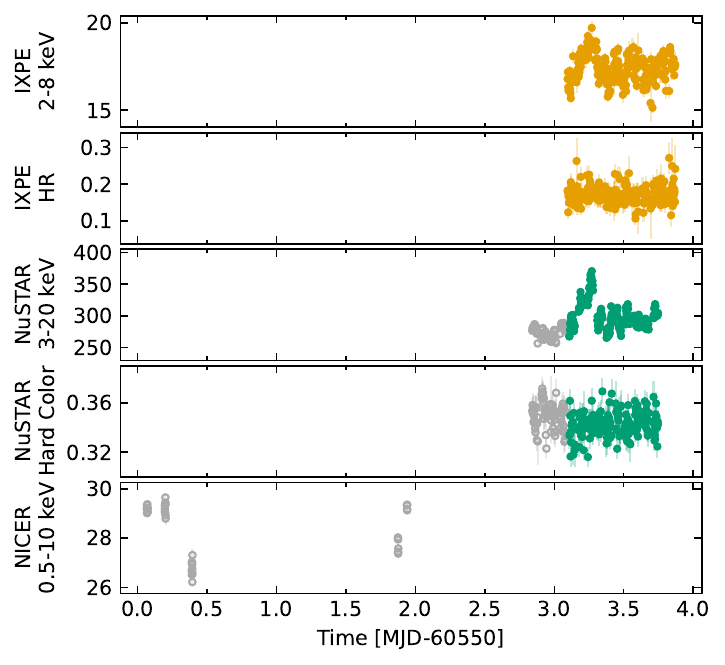}
\caption{Light curves obtained from \ixpe, \nustar, and \nicer. \ixpe HR (5--8 keV/3--5 keV) and \nustar hard color (10--20 keV)/(6--10 keV) are also shown. \ixpe observations are shown in orange, simultaneous \nustar observations are marked in green and non-simultaneous observations are marked in grey.}

\label{fig:all_LC}
\end{figure} 

We performed model-independent polarimetry using the \texttt{pcube} tool. 
The normalized Stokes parameters for the 2--8 keV band are plotted in Figure~\ref{fig:averpol}a and the polarization contours for PD and PA are shown in Figure~\ref{fig:averpol}b. 
Combining the data from all three DUs, we obtain a marginally significant (2$\sigma$) detection of polarization with $\rm PD = 1.4\%\pm0.7\%$ and $\rm PA=-29\degr\pm14\degr$, which corresponds to a $3\sigma$ upper limit for the PD of 3.5\%. 
 
We then analyzed the energy dependence of the polarimetric properties. 
The results are displayed in Figure~\ref{fig:polenergy}. 
The lower panel shows also the MDP99, i.e. the maximum polarization produced by random fluctuations when a true polarization is zero at a confidence level of 99\%.
A polarization signal is detected at the 99\% CL in the 6--7 keV energy bin only with the PD$=8.5\pm 2.9\%$ and PA=$-14\degr\pm 10\degr$, while the PD is clearly below the MDP99 of $\sim3\%$ below 4~keV.
This may suggest a possible energy dependence of the PD, but a higher statistical significance is required to draw firm conclusions. 
We tested alternative binning schemes; however, broader bins did not improve the net significance.

Finally,  no evidence for spectral hardness changes is observed, as seen in Figure~\ref{fig:all_LC}. Thus,  we do not expect large variations in polarization over the observation.

\subsection{Spectroscopic analysis}
\label{sec:spec_analysis}

\nicer observations happened one day before the \ixpe or \nustar observations (see Figure~\ref{fig:all_LC}). Although we attempted a joint spectral fit,  the \nicer spectrum was not compatible with those from  \ixpe or \nustar, likely due to being in a different spectral state, and no satisfactory fit could be obtained. Hence, we used \nustar + \ixpe data to perform the spectroscopic analysis.

We began the spectroscopic analysis with a rather simple model. We used the \texttt{diskbb} model to simulate the emission coming from the accretion disk and the \texttt{comptt} model to simulate the Comptonized emission of the BL/SL region. We chose the spherical geometry of the Comptonizing media. We used the \texttt{tbabs} model \citep{Wilms2000} to account for the absorption of the source emission in the interstellar medium. As neither \ixpe nor \nustar have coverage below 2 keV, we used the HEASARC $N_{\rm H}$ calculator tool based on the survey reported in \citet{HI4PI2016}.

\begin{deluxetable}{lll}
\tablecaption{Best-fitting model parameters of the fits to the \nustar+ \ixpe data.}
\label{table:spectr}
\tablehead{Parameter & Model 1 & Model 2}
\startdata
   \multicolumn{3}{c}{\texttt{tbabs}} \\
$N_{\rm H}$ ($10^{22}$~cm$^{-2}$)\tablenotemark{a} & $[0.28]$ & $[0.28]$ \\
\hline
   \multicolumn{3}{c}{\texttt{diskbb}} \\
$kT_{\rm in}$ (keV) & $0.79\pm0.02$ & $0.90^{+0.04}_{-0.03}$ \\
$\rm norm_{\rm d}$& $400\pm35$ & $240\pm30$  \\
\hline
 \multicolumn{3}{c}{\texttt{comptt}} \\
 $kT_0$ (keV) & $1.11\pm0.02$ & $1.34^{+0.07}_{-0.06}$\\
 $kT_{\rm p}$ (keV) & $3.01\pm0.02$ & $2.96\pm0.02$\\
 $\tau$ & $12.0\pm0.1$ & $13.6\pm0.2$ \\
$\rm norm_{\rm bb}$ & $0.27\pm0.01$ & $0.19\pm0.01$ \\
 \hline
 \multicolumn{3}{c}{\texttt{gauss}} \\
$E$ (keV) & $6.51^{+0.04}_{-0.05}$ & $-$  \\
$\sigma$ (keV) & $0.88^{+0.06}_{-0.05}$ & $-$ \\
$\rm norm_g$ & $0.006\pm0.001$ & $-$ \\
\hline
\multicolumn{3}{c}{\texttt{relxillNS}} \\
$q_{\rm em}$ & $-$ & $[3.0]$ \\
$a$ & $-$ & $[0.0]$ \\
incl (deg) & $-$ & $39^{+3}_{-2}$ \\
$R_{\rm in}$ (ISCO) & $-$ & $3.7^{+1.2}_{-0.8}$ \\
$R_{\rm out}$ (grav. r.)& $-$ & $[400.0]$ \\
$kT_{\rm bb}$ (keV) & $-$ & [=$kT_{\rm p, \texttt{comptt}}$] \\
$\log\xi$ & $-$ & $2.70^{+0.06}_{-0.12}$ \\
$A_{\rm Fe}$ & $-$ & [1.0] \\
$\log N$ & $-$ & [19.0] \\
 $\rm norm_{\rm r}$ ($10^{-3}$) & $-$ & $1.8\pm0.2$ \\
 \hline
\multicolumn{3}{c}{Cross-calibration constants} \\
$C_{\rm DU1-FPMA}$ & $0.874\pm0.002$& $0.873\pm0.002$ \\
$C_{\rm DU2-FPMA}$ & $0.879\pm0.002$& $0.879\pm0.002$ \\
$C_{\rm DU3-FPMA}$ & $0.857\pm0.002$& $0.856\pm0.002$ \\
$C_{\rm FPMB-FPMA}$ & $0.986 \pm 0.001$ & $0.986 \pm 0.001$ \\
\hline
$\chi^2$/d.o.f. & 711/648 & 679/647\\
\hline 
\multicolumn{3}{c}{Photon flux ratios, 2--8 keV} \\
$F_{\rm diskbb}/F_{\rm tot}$ & 0.279 &0.327\\
$F_{\rm comptt}/F_{\rm tot}$ & 0.707 & 0.476 \\
$F_{\rm gauss}/F_{\rm tot}$ &0.014 & $-$ \\
$F_{\rm relxillns}/F_{\rm tot}$ & $-$ & 0.197 \\
$F_{\rm tot}$ (erg~s$^{-1}$~cm$^{-2}$) & $4.0\times 10^{-9}$& $4.0\times 10^{-9}$ \\
\enddata 
\tablecomments{Uncertainties are given at a 68\% CL. Parameters in square brackets were kept frozen during the fit. Flux is unabsorbed.
\tablenotetext{a}{This parameter comes from the HEASARC $N_{\rm H}$ calculator tool based on a HI4PI survey \citep{HI4PI2016}.}}
\end{deluxetable}

\begin{figure*}
\centering
\includegraphics[width=0.45\linewidth]{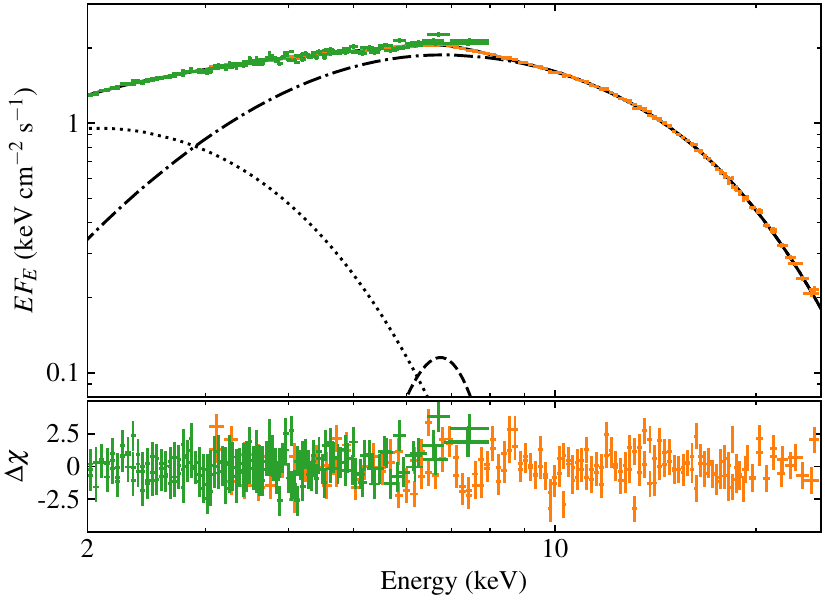} 
\hspace{0.7cm}
\includegraphics[width=0.45\linewidth]{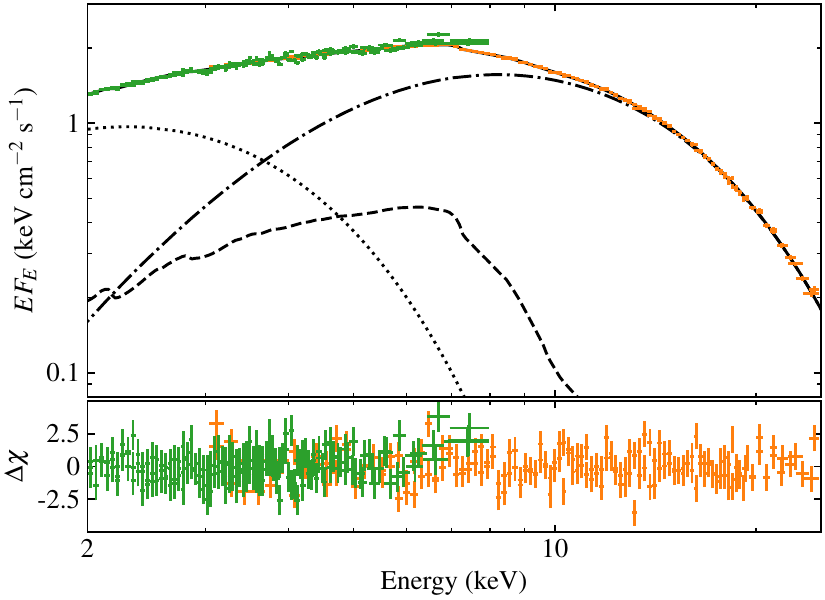}
\caption{Spectral energy distribution of \source. 
The \ixpe and \nustar data are shown with green and orange crosses, respectively. Left panel: fit with the \texttt{tbabs*(diskbb+comptt+gauss)} model. Right panel: fit with the \texttt{tbabs*(diskbb+comptt+relxillNS)} model. 
The lower panels show the fit residuals.
\label{fig:spectrum}}
\end{figure*}

However, the residuals strongly suggested adding a component to account for the reflection of SL/BL emission from the accretion disk. We first added the \texttt{gauss} model to simulate the iron line. We obtained a rather broad gaussian component with $\sigma=0.88^{+0.06}_{-0.05}~\rm keV$ at $E=6.52^{+0.04}_{-0.05}~\rm keV$, which is an expected energy of the K$\alpha$ iron line. The results are presented in Table~\ref{table:spectr}, ``Model 1''  column, and in Figure~\ref{fig:spectrum}, left panel.
We then replaced the \texttt{gauss} component with the \texttt{relxillNS} \citep{2022ApJ...926...13G} model. Results of this fit are presented in Table~\ref{table:spectr}, ``Model 2'' column, and in Figure~\ref{fig:spectrum}, right panel. We note a higher quality of the fit with the \texttt{relxill} model. 

We tied the blackbody temperature to that of the plasma in the \texttt{comptt} model and kept the reflected fraction at $-1.0$ to model only the reflected component without the blackbody. For the \texttt{relxillNS} component, we left the inclination, the inner radius $R_{\rm in}$, the ionization of the accretion disk $\log\xi$, and the normalization of the model to change freely. We fixed the dimensionless spin of the NS at $a=0.0$ as most NSs in LMXBs have $a\leq0.3$ \citep{Galloway2008, Miller2011}. The lack of data in the low-energy range limited our constraint capabilities. Thus, based on previous spectroscopic analyses of the \source \citep{2020ApJ...895...45L}, we fixed the emissivity for the coronal flavor models at $q=3.0$, the density of the accretion disk at $\log N = 19.0$, and the iron abundance at solar level, $A_{\rm Fe}=1.0$.

The resulting inclination of 39\degr\ is lower than the values inferred from previous X-ray observations by \citet{2013A&A...555A..17M} ($\sim$ 60\degr) and \citet{2020ApJ...895...45L} (42\degr--57\degr), but is within the broader range of 27\degr--60\degr\ derived from optical observations by \citet{2006MNRAS.373.1235C}. 
Temperatures of the disk $kT_{\rm in}$ and blackbody emission $kT_{\rm bb}$, as well as the Comptonizing media temperature $kT_{\rm p}$, are in agreement with the previous spectroscopic studies of the source \citep[see, e.g.,][]{2013A&A...555A..17M, 2020ApJ...895...45L, 2024MNRAS.534.2783L}. 
For the inclination obtained from the fit, the disk component normalization corresponds to an inner radius $R_{\rm in, \texttt{diskbb}}\approx$10--12~km, depending on the model. 
According to the correlation between the apparent and true inner radii of the disk given in, for example,  \citet{1998PASJ...50..667K}, the true inner radius of the disk is only a few km larger than the standard NS radius. 
Inner radius of the accretion disk from the \texttt{relxillNS} model, however, is constrained at $R_{\rm in, \texttt{relxillNS}}=3.7^{+1.2}_{-0.8}~R_{\rm ISCO}\approx45$~km.

\subsection{Spectro-polarimetric analysis}\label{sec:spectro}

To conduct the spectro-polarimetric analysis, we used the spectral model presented in Section~\ref{sec:spec_analysis} and added the polarimetric model \texttt{polconst}. 
We again fitted \nustar+ \ixpe data and obtained the values of the overall $\rm PD = 1.0\%\pm0.5\%$ and $\rm PA = -25\degr\pm15\degr$ (1$\sigma$ CL). 
It is, however, important to note that we can only obtain an upper limit of $2.4\%$ on the PD at $3\sigma$ CL, which means that we do not have a strong detection of polarization and the PA is not  well constrained. 
We then attempted to use the \texttt{pollin} polarimetric model and fixed the slope on PA at 0. 
We obtained $\rm PD_{1 keV} = -1.3\%\pm1.2\%$, $\rm PD_{slope}=0.9\pm0.5\%\,\mbox{keV}^{-1}$ and a $\rm PA=-29\degr\pm11\degr$ (errors at $1\sigma$ CL).\footnote{The negative value of the PD at 1~keV and a positive PD slope imply a change of the PA  by 90\degr\ between 1~keV and higher energies, i.e. the actual value of the PA at 1~keV is 63\degr.} We performed an F-test to check if the \texttt{pollin} model improved the fitting results significantly, and obtained a probability of 0.082. Hence, we did not see any significant improvement of the fit with \texttt{pollin}.

\begin{deluxetable}{lll}
\tablecaption{Spectro-polarimetric constraints on the polarization of different spectral components from the \nustar+ \ixpe data.}
\label{tab:spectropol}
\tablehead{Component & PD (\%) & PA (deg)}  
\startdata 
overall & $1.0\pm0.5$ & $-25\pm15$ \\
\hline 
\texttt{diskbb} & [1] & $54^{+36}_{-39}$ \\
\texttt{comptt} & [1] & $-35^{+60}_{-62}$ \\
\texttt{relxillNS} & $5.6^{+4.0}_{-2.8}$ & $-27^{+21}_{-24}$ \\
\hline  
\texttt{diskbb} & $1.5\pm1.1$ & $43^{+43}_{-26}$ \\
\texttt{comptt} & [1] & $-31^{+32}_{-34}$ \\
\texttt{relxillNS} & [5] & $-35^{+25}_{-22}$ \\
\hline
\texttt{diskbb} & $0.9\pm0.5$ & $-25\pm15$ \\
\texttt{comptt} & $=\textrm{PD}_{\texttt{diskbb}}$ & $=\textrm{PA}_{\texttt{diskbb}}$  \\
\texttt{gauss} & [0] & [0] \\
\enddata
\tablecomments{Errors are given at 68\% CL. Parameters in square brackets were kept frozen during the fit.}
\end{deluxetable}

We proceeded with a study of the polarization of each component independently. 
As we only have a marginal detection of polarization, such a study would not be possible without some artificial constraints. 
Results of this analysis are presented in Table~\ref{tab:spectropol}. 
First, we fixed the polarization of the \texttt{diskbb} and \texttt{comptt} components to be at $\rm PD=1\%$, following the expected polarization from these components \citep{Loktev22, Farinelli24, Bobrikova25}, and let the polarization of the \texttt{relxillNS} component to vary freely. 
We obtained a $\rm PD_{\texttt{relxillNS}}=5.6^{+4.0}_{-2.8}\%$ at $1\sigma$ CL with the $3\sigma$ upper limit of 15\%. 
At $1\sigma$ CL, we also obtained constraints on PAs of different components and saw that disk emission seems to be polarized almost orthogonal to the Comptonized component and the reflection. 
The difference in PA of $\sim90\degr$ between \texttt{diskbb} and other components is consistent with the results of the fit with \texttt{pollin} model which show a flip of the PA by 90\degr\ in the \ixpe  range.   
We note, however, that the error bars on PA are large for all the components.

Then, we fixed the \texttt{relxillNS} component polarization at 5\% based on our previous results and allowed the \texttt{diskbb} polarization to vary, obtaining a $\rm PD_{\texttt{diskbb}}=1.5\%\pm1.1\%$ at $1\sigma$ CL with the $3\sigma$ upper limit of 5\%. 
We saw again $\rm PA_{\texttt{relxillNS}}$ almost coinciding with $\rm PA_{\texttt{comptt}}$ within $1\sigma$ errors, while $\rm PA_{\texttt{diskbb}}$ differs by $\approx75\degr$. With this test, we also confirmed that fixing and thawing various components did not have a strong influence on the polarimetric picture. 
Unfortunately, the statistics did not allow us to perform any further quantitative investigations of the spectro-polarimetric properties using Model 2. 

Lastly, we came back to Model 1 from Table~\ref{table:spectr}. As the fluorescent iron line is supposed to be unpolarized, we fixed PD and PA for the \texttt{gauss} component to zero. As we aimed at studying the overall polarization of the source emission, we tied the polarimetric properties of the \texttt{comptt} component to the ones of the \texttt{diskbb} component. We obtained a $\rm PD = 0.9\%\pm0.5\%$ and $\rm PA=-25\degr\pm15\degr$ at $1\sigma$ CL with the $3\sigma$ upper limit on the PD of 2.4\%. These results are in alignment with the ones obtained from the spectro-polarimetric analysis using Model~2.

\section{Discussion and summary}
\label{sec:discussion}

We used \ixpe, \nustar, and \nicer data to investigate the X-ray spectro-polarimetric properties of \source. 
With the model-independent analysis, we obtained a marginal detection of polarization with a PD of $1.4\%\pm0.7\%$ and PA of $-29\degr\pm14\degr$ with the corresponding $3\sigma$ upper limit on the PD of 3.5\%.
These results, together with the presence of reflection features and a low estimated inclination of 39\degr, are broadly consistent with the results of other atolls. 
The spectro-polarimetric analysis  yielded a $\rm PD = 1.0\%\pm0.5\%$ and $\rm PA = -25\degr\pm15\degr$ with the corresponding $3\sigma$ upper limit on the PD of 2.4\%, which is consistent with the values obtained from the \texttt{pcube} polarimetric analysis. 

Among other atoll sources, these results are similar to \mbox{GS~1826$-$238} \citep{Capitanio23}, \mbox{Ser~X-1} \citep{SerX1}, \mbox{GX~3+1}  \citep{Gnarini24} and \mbox{GX~9+1} \citep{Tarana_2025}, for which only upper limits could be obtained at $3\sigma$ level. The case of \source appears to be consistent in particular with that of \mbox{Ser~X-1}, \mbox{GX~3+1}, and \mbox{GX~9+1}, which, due to their low inclination, showed no polarization detection despite exhibiting reflection features. This suggests \source may have a similar system geometry and configuration of the Comptonization region.

For future work, longer observations with \ixpe should enable tighter constraints on the polarization parameters of \source. In addition to the increased exposure time, applying an event-based maximum likelihood estimation analysis, as described in \citet{Herman2021a,Herman2021b,Herman2024} and implemented as a Bayesian Nested Sampling (BNS) method by \citet{Ravi2026}, can also enhance the detection significance. A longer exposure and stronger polarization signal would shed further light onto the source’s behavior.

\section*{Acknowledgments} 
The Imaging X-ray Polarimetry Explorer (IXPE) is a joint US and Italian mission.  The US contribution is supported by the National Aeronautics and Space Administration (NASA) and led and managed by its Marshall Space Flight Center (MSFC), with industry partner Ball Aerospace (contract NNM15AA18C). 
This research used data products provided by the IXPE Team (MSFC, SSDC, INAF, and INFN) and distributed with additional software tools by the High-Energy Astrophysics Science Archive Research Center (HEASARC), at NASA Goddard Space Flight Center (GSFC). 
This work was partially supported by the Research Council of Finland Centre of Excellence in Neutron-Star Physics (grant 374064).
MADT acknowledges support from the EDUFI Fellowship and the Johannes Andersen Student Programme at the Nordic Optical Telescope. 
AB was supported by the  Finnish Cultural Foundation (grants 002200175 and 00240328).  
AG, FU, SB, FC, SF, GM, and AT  acknowledge financial support from the Italian Space Agency (Agenzia Spaziale Italiana, ASI) through the contract ASI-INAF-2022-19-HH.0 and by the Istituto Nazionale di Astrofisica (INAF) through the grant 1.05.23.05.06: ``Spin and Geometry in accreting X-ray binaries: The first multi frequency spectro-polarimetric campaign''. The Italian contribution was also supported by ASI through contract ASI-OHBI-2022-13-I.0, agreement and ASI-INFN-2017.13-H0, and its Space Science Data Center (SSDC) with agreements ASI-INAF-2022-14-HH.0 and ASI-INFN 2021-43-HH.0, and by the Istituto Nazionale di Fisica Nucleare (INFN) in Italy. 
AG was supported by an appointment to the NASA Postdoctoral Program at the Marshall Space Flight Center (MSFC), administered by Oak Ridge Associated Universities under contract with NASA.
MN is a Fonds de Recherche du Quebec – Nature et Technologies (FRQNT) postdoctoral fellow. 
AS acknowledges the support of the Jenny and Antti Wihuri Foundation (grant no. 00240331).


%

\vspace{5mm}
\facilities{\ixpe, \nicer, \nustar}





\bibliography{ixpe_publ}{}

@ARTICLE{Wilms2000,
       author = {{Wilms}, J. and {Allen}, A. and {McCray}, R.},
        title = "{On the Absorption of X-Rays in the Interstellar Medium}",
      journal = {\apj},
         year = 2000,
        month = oct,
       volume = {542},
       number = {2},
        pages = {914-924},
          doi = {10.1086/317016},
archivePrefix = {arXiv},
       eprint = {astro-ph/0008425},
 primaryClass = {astro-ph},
       adsurl = {https://ui.adsabs.harvard.edu/abs/2000ApJ...542..914W}
}

@ARTICLE{DiMarco2025,
       author = {{Di Marco}, Alessandro and {La Monaca}, Fabio and {Bobrikova}, Anna and {Stella}, Luigi and {Papitto}, Alessandro and {Poutanen}, Juri and {Baglio}, Maria Cristina and {Bachetti}, Matteo and {Loktev}, Vladislav and {Pilia}, Maura and {Rogantini}, Daniele},
        title = "{X-Ray Dips and Polarization Angle Swings in GX 13+1}",
      journal = {\apjl},
         year = 2025,
        month = feb,
       volume = {979},
       number = {2},
          eid = {L47},
        pages = {L47},
          doi = {10.3847/2041-8213/ada7f8},
archivePrefix = {arXiv},
       eprint = {2501.05511},
 primaryClass = {astro-ph.HE},
       adsurl = {https://ui.adsabs.harvard.edu/abs/2025ApJ...979L..47D}
}

@ARTICLE{Suleimanov2006,
       author = {{Suleimanov}, Valery and {Poutanen}, Juri},
        title = "{Spectra of the spreading layers on the neutron star surface and constraints on the neutron star equation of state}",
      journal = {\mnras},
         year = 2006,
        month = jul,
       volume = {369},
       number = {4},
        pages = {2036-2048},
          doi = {10.1111/j.1365-2966.2006.10454.x},
archivePrefix = {arXiv},
       eprint = {astro-ph/0601689},
 primaryClass = {astro-ph},
       adsurl = {https://ui.adsabs.harvard.edu/abs/2006MNRAS.369.2036S}
}

@ARTICLE{1993SSRv...62..223L,
       author = {{Lewin}, Walter H.~G. and {van Paradijs}, Jan and {Taam}, Ronald E.},
        title = "{X-Ray Bursts}",
      journal = {\ssr},
         year = 1993,
        month = sep,
       volume = {62},
       number = {3-4},
        pages = {223-389},
          doi = {10.1007/BF00196124},
       adsurl = {https://ui.adsabs.harvard.edu/abs/1993SSRv...62..223L}
}

@ARTICLE{Matt1993,
       author = {{Matt}, Giorgio},
        title = "{X-ray polarization properties of a centrally illuminated accretion disc.}",
      journal = {\mnras},
         year = 1993,
        month = feb,
       volume = {260},
        pages = {663-674},
          doi = {10.1093/mnras/260.3.663},
          adsurl = {https://ui.adsabs.harvard.edu/abs/1993MNRAS.260..663M}
}

@ARTICLE{Poutanen96,
       author = {{Poutanen}, Juri and {Nagendra}, K.~N. and {Svensson}, Roland},
        title = "{Green's matrix for Compton reflection of polarized radiation from cold matter}",
      journal = {\mnras},
         year = 1996,
        month = dec,
       volume = {283},
       number = {3},
        pages = {892-904},
          doi = {10.1093/mnras/283.3.892},
       adsurl = {https://ui.adsabs.harvard.edu/abs/1996MNRAS.283..892P}
}

@ARTICLE{1989A&A...225...79H,
       author = {{Hasinger}, G. and {van der Klis}, M.},
        title = "{Two patterns of correlated X-ray timing and spectral behaviour in low-mass X-ray binaries.}",
      journal = {\aap},
         year = 1989,
        month = nov,
       volume = {225},
        pages = {79-96},
       adsurl = {https://ui.adsabs.harvard.edu/abs/1989A&A...225...79H}
}

@INPROCEEDINGS{Bahramian2024,
       author = {{Bahramian}, Arash and {Degenaar}, Nathalie},
        title = "{Low-Mass X-ray Binaries}",
    booktitle = {Handbook of X-ray and Gamma-ray Astrophysics},
publisher = {Springer},
 address = {Singapore}, 
    editor = {{Bambi}, C. and {Santangelo}, A.},
year = 2024, 
        pages = {3657-3718},
          doi = {10.1007/978-981-19-6960-7_94},
archivePrefix = {arXiv},
       eprint = {2206.10053},
 primaryClass = {astro-ph.HE},
       adsurl = {https://ui.adsabs.harvard.edu/abs/2023hxga.book..120B}
}

@ARTICLE{ShakuraSunyaev73,
       author = {{Shakura}, N.~I. and {Sunyaev}, R.~A.},
        title = "{Black holes in binary systems. Observational appearance.}",
      journal = {\aap},
         year = 1973,
        month = jan,
       volume = {24},
        pages = {337-355},
       adsurl = {https://ui.adsabs.harvard.edu/abs/1973A&A....24..337S}
}

@ARTICLE{HI4PI2016,
       author = {{HI4PI Collaboration} and {Ben Bekhti}, N. and {Fl{\"o}er}, L. and {Keller}, R. and {Kerp}, J. and {Lenz}, D. and {Winkel}, B. and {Bailin}, J. and {Calabretta}, M.~R. and {Dedes}, L. and {Ford}, H.~A. and {Gibson}, B.~K. and {Haud}, U. and {Janowiecki}, S. and {Kalberla}, P.~M.~W. and {Lockman}, F.~J. and {McClure-Griffiths}, N.~M. and {Murphy}, T. and {Nakanishi}, H. and {Pisano}, D.~J. and {Staveley-Smith}, L.},
        title = "{HI4PI: A full-sky H I survey based on EBHIS and GASS}",
      journal = {\aap},
         year = 2016,
        month = oct,
       volume = {594},
          eid = {A116},
        pages = {A116},
          doi = {10.1051/0004-6361/201629178},
archivePrefix = {arXiv},
       eprint = {1610.06175},
 primaryClass = {astro-ph.GA},
       adsurl = {https://ui.adsabs.harvard.edu/abs/2016A&A...594A.116H}
}

@ARTICLE{Dovciak2008,
       author = {{Dov{\v{c}}iak}, M. and {Muleri}, F. and {Goosmann}, R.~W. and {Karas}, V. and {Matt}, G.},
        title = "{Thermal disc emission from a rotating black hole: X-ray polarization signatures}",
      journal = {\mnras},
         year = 2008,
        month = nov,
       volume = {391},
       number = {1},
        pages = {32-38},
          doi = {10.1111/j.1365-2966.2008.13872.x},
archivePrefix = {arXiv},
       eprint = {0809.0418},
 primaryClass = {astro-ph},
       adsurl = {https://ui.adsabs.harvard.edu/abs/2008MNRAS.391...32D}
}

@ARTICLE{SerX1, 
       author = {{Ursini}, F. and {Gnarini}, A. and {Bianchi}, S. and {Bobrikova}, A. and {Capitanio}, F. and {Cocchi}, M. and {Fabiani}, S. and {Farinelli}, R. and {Kaaret}, P. and {Matt}, G. and {Ng}, M. and {Poutanen}, J. and {Ravi}, S. and {Tarana}, A.},
        title = "{X-ray spectropolarimetry of the bright atoll Serpens X-1}",
      journal = {\aap},
         year = 2024,
        month = oct,
       volume = {690},
          eid = {A200},
        pages = {A200},
          doi = {10.1051/0004-6361/202451584},
archivePrefix = {arXiv},
       eprint = {2408.16713},
 primaryClass = {astro-ph.HE},
       adsurl = {https://ui.adsabs.harvard.edu/abs/2024A&A...690A.200U}
}

@ARTICLE{2021APh...13302628B,
       author = {{Baldini}, L. and {Barbanera}, M. and {Bellazzini}, R. and {Bonino}, R. and {Borotto}, F. and {Brez}, A. and {Caporale}, C. and {Cardelli}, C. and {Castellano}, S. and {Ceccanti}, M. and {Citraro}, S. and {Di Lalla}, N. and {Latronico}, L. and {Lucchesi}, L. and {Magazz{\`u}}, C. and {Magazz{\`u}}, G. and {Maldera}, S. and {Manfreda}, A. and {Marengo}, M. and {Marrocchesi}, A. and {Mereu}, P. and {Minuti}, M. and {Mosti}, F. and {Nasimi}, H. and {Nuti}, A. and {Oppedisano}, C. and {Orsini}, L. and {Pesce-Rollins}, M. and {Pinchera}, M. and {Profeti}, A. and {Sgr{\`o}}, C. and {Spandre}, G. and {Tardiola}, M. and {Zanetti}, D. and {Amici}, F. and {Andersson}, H. and {Attin{\`a}}, P. and {Bachetti}, M. and {Baumgartner}, W. and {Brienza}, D. and {Carpentiero}, R. and {Castronuovo}, M. and {Cavalli}, L. and {Cavazzuti}, E. and {Centrone}, M. and {Costa}, E. and {D'Alba}, E. and {D'Amico}, F. and {Del Monte}, E. and {Di Cosimo}, S. and {Di Marco}, A. and {Di Persio}, G. and {Donnarumma}, I. and {Evangelista}, Y. and {Fabiani}, S. and {Ferrazzoli}, R. and {Kitaguchi}, T. and {La Monaca}, F. and {Lefevre}, C. and {Loffredo}, P. and {Lorenzi}, P. and {Mangraviti}, E. and {Matt}, G. and {Meilahti}, T. and {Morbidini}, A. and {Muleri}, F. and {Nakano}, T. and {Negri}, B. and {Nenonen}, S. and {O'Dell}, S.~L. and {Perri}, M. and {Piazzolla}, R. and {Pieraccini}, S. and {Pilia}, M. and {Puccetti}, S. and {Ramsey}, B.~D. and {Rankin}, J. and {Ratheesh}, A. and {Rubini}, A. and {Santoli}, F. and {Sarra}, P. and {Scalise}, E. and {Sciortino}, A. and {Soffitta}, P. and {Tamagawa}, T. and {Tennant}, A.~F. and {Tobia}, A. and {Trois}, A. and {Uchiyama}, K. and {Vimercati}, M. and {Weisskopf}, M.~C. and {Xie}, F. and {Zanetti}, F. and {Zhou}, Y.},
        title = "{Design, construction, and test of the Gas Pixel Detectors for the IXPE mission}",
      journal = {Astroparticle Physics},
         year = 2021,
        month = dec,
       volume = {133},
          eid = {102628},
        pages = {102628},
          doi = {10.1016/j.astropartphys.2021.102628},
archivePrefix = {arXiv},
       eprint = {2107.05496},
 primaryClass = {astro-ph.IM},
       adsurl = {https://ui.adsabs.harvard.edu/abs/2021APh...13302628B}
}

@ARTICLE{2021AJ....162..208S,
       author = {{Soffitta}, Paolo and {Baldini}, Luca and {Bellazzini}, Ronaldo and {Costa}, Enrico and {Latronico}, Luca and {Muleri}, Fabio and {Del Monte}, Ettore and {Fabiani}, Sergio and {Minuti}, Massimo and {Pinchera}, Michele and {Sgro'}, Carmelo and {Spandre}, Gloria and {Trois}, Alessio and {Amici}, Fabrizio and {Andersson}, Hans and {Attina'}, Primo and {Bachetti}, Matteo and {Barbanera}, Mattia and {Borotto}, Fabio and {Brez}, Alessandro and {Brienza}, Daniele and {Caporale}, Ciro and {Cardelli}, Claudia and {Carpentiero}, Rita and {Castellano}, Simone and {Castronuovo}, Marco and {Cavalli}, Luca and {Cavazzuti}, Elisabetta and {Ceccanti}, Marco and {Centrone}, Mauro and {Ciprini}, Stefano and {Citraro}, Saverio and {D'Amico}, Fabio and {D'Alba}, Elisa and {Di Cosimo}, Sergio and {Di Lalla}, Niccolo' and {Di Marco}, Alessandro and {Di Persio}, Giuseppe and {Donnarumma}, Immacolata and {Evangelista}, Yuri and {Ferrazzoli}, Riccardo and {Hayato}, Asami and {Kitaguchi}, Takao and {La Monaca}, Fabio and {Lefevre}, Carlo and {Loffredo}, Pasqualino and {Lorenzi}, Paolo and {Lucchesi}, Leonardo and {Magazzu}, Carlo and {Maldera}, Simone and {Manfreda}, Alberto and {Mangraviti}, Elio and {Marengo}, Marco and {Matt}, Giorgio and {Mereu}, Paolo and {Morbidini}, Alfredo and {Mosti}, Federico and {Nakano}, Toshio and {Nasimi}, Hikmat and {Negri}, Barbara and {Nenonen}, Seppo and {Nuti}, Alessio and {Orsini}, Leonardo and {Perri}, Matteo and {Pesce-Rollins}, Melissa and {Piazzolla}, Raffaele and {Pilia}, Maura and {Profeti}, Alessandro and {Puccetti}, Simonetta and {Rankin}, John and {Ratheesh}, Ajay and {Rubini}, Alda and {Santoli}, Francesco and {Sarra}, Paolo and {Scalise}, Emanuele and {Sciortino}, Andrea and {Tamagawa}, Toru and {Tardiola}, Marcello and {Tobia}, Antonino and {Vimercati}, Marco and {Xie}, Fei},
        title = "{The Instrument of the Imaging X-Ray Polarimetry Explorer}",
      journal = {\aj},
         year = 2021,
        month = nov,
       volume = {162},
       number = {5},
          eid = {208},
        pages = {208},
          doi = {10.3847/1538-3881/ac19b0},
archivePrefix = {arXiv},
       eprint = {2108.00284},
 primaryClass = {astro-ph.IM},
       adsurl = {https://ui.adsabs.harvard.edu/abs/2021AJ....162..208S}
}

@ARTICLE{Weisskopf2022,
       author = {{Weisskopf}, Martin C. and {Soffitta}, Paolo and {Baldini}, Luca and {Ramsey}, Brian D. and {O'Dell}, Stephen L. and {Romani}, Roger W. and {Matt}, Giorgio and {Deininger}, William D. and {Baumgartner}, Wayne H. and {Bellazzini}, Ronaldo and {Costa}, Enrico and {Kolodziejczak}, Jeffery J. and {Latronico}, Luca and {Marshall}, Herman L. and {Muleri}, Fabio and {Bongiorno}, Stephen D. and {Tennant}, Allyn and {Bucciantini}, Niccolo and {Dovciak}, Michal and {Marin}, Frederic and {Marscher}, Alan and {Poutanen}, Juri and {Slane}, Pat and {Turolla}, Roberto and {Kalinowski}, William and {Di Marco}, Alessandro and {Fabiani}, Sergio and {Minuti}, Massimo and {La Monaca}, Fabio and {Pinchera}, Michele and {Rankin}, John and {Sgro'}, Carmelo and {Trois}, Alessio and {Xie}, Fei and {Alexander}, Cheryl and {Allen}, D. Zachery and {Amici}, Fabrizio and {Andersen}, Jason and {Antonelli}, Angelo and {Antoniak}, Spencer and {Attina'}, Primo and {Barbanera}, Mattia and {Bachetti}, Matteo and {Baggett}, Randy M. and {Bladt}, Jeff and {Brez}, Alessandro and {Bonino}, Raffaella and {Boree}, Christopher and {Borotto}, Fabio and {Breeding}, Shawn and {Brienza}, Daniele and {Bygott}, H. Kyle and {Caporale}, Ciro and {Cardelli}, Claudia and {Carpentiero}, Rita and {Castellano}, Simone and {Castronuovo}, Marco and {Cavalli}, Luca and {Cavazzuti}, Elisabetta and {Ceccanti}, Marco and {Centrone}, Mauro and {Citraro}, Saverio and {D' Amico}, Fabio and {D'Alba}, Elisa and {Di Gesu}, Laura and {Del Monte}, Ettore and {Dietz}, Kurtis L. and {Di Lalla}, Niccolo' and {Di Persio}, Giuseppe and {Dolan}, David and {Donnarumma}, Immacolata and {Evangelista}, Yuri and {Ferrant}, Kevin and {Ferrazzoli}, Riccardo and {Ferrie}, MacKenzie and {Footdale}, Joseph and {Forsyth}, Brent and {Foster}, Michelle and {Garelick}, Benjamin and {Gunji}, Shuichi and {Gurnee}, Eli and {Head}, Michael and {Hibbard}, Grant and {Johnson}, Samantha and {Kelly}, Erik and {Kilaru}, Kiranmayee and {Lefevre}, Carlo and {Le Roy}, Shelley and {Loffredo}, Pasqualino and {Lorenzi}, Paolo and {Lucchesi}, Leonardo and {Maddox}, Tyler and {Magazzu}, Guido and {Maldera}, Simone and {Manfreda}, Alberto and {Mangraviti}, Elio and {Marengo}, Marco and {Marrocchesi}, Alessandra and {Massaro}, Francesco and {Mauger}, David and {McCracken}, Jeffrey and {McEachen}, Michael and {Mize}, Rondal and {Mereu}, Paolo and {Mitchell}, Scott and {Mitsuishi}, Ikuyuki and {Morbidini}, Alfredo and {Mosti}, Federico and {Nasimi}, Hikmat and {Negri}, Barbara and {Negro}, Michela and {Nguyen}, Toan and {Nitschke}, Isaac and {Nuti}, Alessio and {Onizuka}, Mitch and {Oppedisano}, Chiara and {Orsini}, Leonardo and {Osborne}, Darren and {Pacheco}, Richard and {Paggi}, Alessandro and {Painter}, Will and {Pavelitz}, Steven D. and {Pentz}, Christina and {Piazzolla}, Raffaele and {Perri}, Matteo and {Pesce-Rollins}, Melissa and {Peterson}, Colin and {Pilia}, Maura and {Profeti}, Alessandro and {Puccetti}, Simonetta and {Ranganathan}, Jaganathan and {Ratheesh}, Ajay and {Reedy}, Lee and {Root}, Noah and {Rubini}, Alda and {Ruswick}, Stephanie and {Sanchez}, Javier and {Sarra}, Paolo and {Santoli}, Francesco and {Scalise}, Emanuele and {Sciortino}, Andrea and {Schroeder}, Christopher and {Seek}, Tim and {Sosdian}, Kalie and {Spandre}, Gloria and {Speegle}, Chet O. and {Tamagawa}, Toru and {Tardiola}, Marcello and {Tobia}, Antonino and {Thomas}, Nicholas E. and {Valerie}, Robert and {Vimercati}, Marco and {Walden}, Amy L. and {Weddendorf}, Bruce and {Wedmore}, Jeffrey and {Welch}, David and {Zanetti}, Davide and {Zanetti}, Francesco},
        title = "{The Imaging X-Ray Polarimetry Explorer (IXPE): Pre-Launch}",
      journal = {JATIS},
         year = 2022,
        month = apr,
       volume = {8},
       number = {2},
       pages = {026002},
          doi = {10.1117/1.JATIS.8.2.026002},
archivePrefix = {arXiv},
       eprint = {2112.01269},
 primaryClass = {astro-ph.IM},
       adsurl = {https://ui.adsabs.harvard.edu/abs/2021arXiv211201269W}
}

@ARTICLE{Baldini2022,
       author = {{Baldini}, L. and {Bucciantini}, N. and {Di Lalla}, N. and {Ehlert}, S.~R. and {Manfreda}, A. and {Omodei}, N. and {Pesce-Rollins}, M. and {Sgr{\`o}}, C.},
        title = "{ixpeobssim: a Simulation and Analysis Framework for the Imaging X-ray Polarimetry Explorer}",
      journal = {SoftwareX},
         year = 2022,
        month = mar,
         volume= {19},
          eid = {101194},
          pages = {101194},
          doi = {10.1016/j.softx.2022.101194},
archivePrefix = {arXiv},
       eprint = {2203.06384},
 primaryClass = {astro-ph.IM},
       adsurl = {https://ui.adsabs.harvard.edu/abs/2022arXiv220306384B}
}

@ARTICLE{2015APh....68...45K,
       author = {{Kislat}, F. and {Clark}, B. and {Beilicke}, M. and {Krawczynski}, H.},
        title = "{Analyzing the data from X-ray polarimeters with Stokes parameters}",
      journal = {Astroparticle Physics},
         year = 2015,
        month = aug,
       volume = {68},
        pages = {45-51},
          doi = {10.1016/j.astropartphys.2015.02.007},
archivePrefix = {arXiv},
       eprint = {1409.6214},
 primaryClass = {astro-ph.IM},
       adsurl = {https://ui.adsabs.harvard.edu/abs/2015APh....68...45K}
}

@ARTICLE{Gnarini22,
       author = {{Gnarini}, A. and {Ursini}, F. and {Matt}, G. and {Bianchi}, S. and {Capitanio}, F. and {Cocchi}, M. and {Farinelli}, R. and {Zhang}, W.},
        title = "{Polarization properties of weakly magnetized neutron stars in low-mass X-ray binaries}",
      journal = {\mnras},
         year = 2022,
        month = aug,
       volume = {514},
       number = {2},
        pages = {2561-2567},
          doi = {10.1093/mnras/stac1523},
archivePrefix = {arXiv},
       eprint = {2206.00749},
 primaryClass = {astro-ph.HE},
       adsurl = {https://ui.adsabs.harvard.edu/abs/2022MNRAS.514.2561G}
}

@ARTICLE{Gnarini24,
       author = {{Gnarini}, Andrea and {Farinelli}, Ruben and {Ursini}, Francesco and {Bianchi}, Stefano and {Capitanio}, Fiamma and {Matt}, Giorgio and {Ng}, Mason and {Tarana}, Antonella and {Bobrikova}, Anna and {Cocchi}, Massimo and {Fabiani}, Sergio and {Kaaret}, Philip and {Poutanen}, Juri and {Ravi}, Swati},
        title = "{First spectropolarimetric observation of the neutron star low-mass X-ray binary GX 3+1}",
      journal = {\aap},
         year = 2024,
        month = dec,
       volume = {692},
          eid = {A123},
        pages = {A123},
          doi = {10.1051/0004-6361/202452642},
archivePrefix = {arXiv},
       eprint = {2411.10353},
 primaryClass = {astro-ph.HE},
       adsurl = {https://ui.adsabs.harvard.edu/abs/2024A&A...692A.123G}
}

@ARTICLE{2001ApJ...547..355P,
       author = {{Popham}, Robert and {Sunyaev}, Rashid},
        title = "{Accretion Disk Boundary Layers around Neutron Stars: X-Ray Production in Low-Mass X-Ray Binaries}",
      journal = {\apj},
         year = 2001,
        month = jan,
       volume = {547},
       number = {1},
        pages = {355-383},
          doi = {10.1086/318336},
archivePrefix = {arXiv},
       eprint = {astro-ph/0004017},
 primaryClass = {astro-ph},
       adsurl = {https://ui.adsabs.harvard.edu/abs/2001ApJ...547..355P}
}

@ARTICLE{IS1999,
       author = {{Inogamov}, N.~A. and {Sunyaev}, R.~A.},
        title = "{Spread of matter over a neutron-star surface during disk accretion}",
      journal = {Astronomy Letters},
         year = 1999,
        month = may,
       volume = {25},
       number = {5},
        pages = {269-293},
          doi = {10.48550/arXiv.astro-ph/9904333},
archivePrefix = {arXiv},
       eprint = {astro-ph/9904333},
 primaryClass = {astro-ph},
       adsurl = {https://ui.adsabs.harvard.edu/abs/1999AstL...25..269I}
}

@ARTICLE{DiMarco23,
       author = {{Di Marco}, Alessandro and {La Monaca}, Fabio and {Poutanen}, Juri and {Russell}, Thomas D. and {Anitra}, Alessio and {Farinelli}, Ruben and {Mastroserio}, Guglielmo and {Muleri}, Fabio and {Xie}, Fei and {Bachetti}, Matteo and {Burderi}, Luciano and {Carotenuto}, Francesco and {Del Santo}, Melania and {Di Salvo}, Tiziana and {Dov{\v{c}}iak}, Michal and {Gnarini}, Andrea and {Iaria}, Rosario and {Kajava}, Jari J.~E. and {Liu}, Kuan and {Middei}, Riccardo and {O'Dell}, Stephen L. and {Pilia}, Maura and {Rankin}, John and {Sanna}, Andrea and {Eijnden}, Jakob van den and {Weisskopf}, Martin C. and {Bobrikova}, Anna and {Capitanio}, Fiamma and {Costa}, Enrico and {Kaaret}, Philip and {Marino}, Alessio and {Soffitta}, Paolo and {Ursini}, Francesco and {Ambrosino}, Filippo and {Cocchi}, Massimo and {Fabiani}, Sergio and {Marshall}, Herman L. and {Matt}, Giorgio and {Motta}, Sara Elisa and {Papitto}, Alessandro and {Stella}, Luigi and {Tarana}, Antonella and {Zane}, Silvia and {Agudo}, Iv{\'a}n and {Antonelli}, Lucio A. and {Baldini}, Luca and {Baumgartner}, Wayne H. and {Bellazzini}, Ronaldo and {Bianchi}, Stefano and {Bongiorno}, Stephen D. and {Bonino}, Raffaella and {Brez}, Alessandro and {Bucciantini}, Niccol{\`o} and {Castellano}, Simone and {Cavazzuti}, Elisabetta and {Chen}, Chien-Ting and {Ciprini}, Stefano and {De Rosa}, Alessandra and {Del Monte}, Ettore and {Di Gesu}, Laura and {Di Lalla}, Niccol{\`o} and {Donnarumma}, Immacolata and {Doroshenko}, Victor and {Ehlert}, Steven R. and {Enoto}, Teruaki and {Evangelista}, Yuri and {Ferrazzoli}, Riccardo and {Garcia}, Javier A. and {Gunji}, Shuichi and {Hayashida}, Kiyoshi and {Heyl}, Jeremy and {Iwakiri}, Wataru and {Jorstad}, Svetlana G. and {Karas}, Vladimir and {Kislat}, Fabian and {Kitaguchi}, Takao and {Kolodziejczak}, Jeffery J. and {Krawczynski}, Henric and {Latronico}, Luca and {Liodakis}, Ioannis and {Maldera}, Simone and {Manfreda}, Alberto and {Marin}, Fr{\'e}d{\'e}ric and {Marinucci}, Andrea and {Marscher}, Alan P. and {Massaro}, Francesco and {Mitsuishi}, Ikuyuki and {Mizuno}, Tsunefumi and {Negro}, Michela and {Ng}, Chi-Yung and {Omodei}, Nicola and {Oppedisano}, Chiara and {Pavlov}, George G. and {Peirson}, Abel L. and {Perri}, Matteo and {Pesce-Rollins}, Melissa and {Petrucci}, Pierre-Olivier and {Possenti}, Andrea and {Puccetti}, Simonetta and {Ramsey}, Brian D. and {Ratheesh}, Ajay and {Roberts}, Oliver J. and {Romani}, Roger W. and {Sgr{\`o}}, Carmelo and {Slane}, Patrick and {Spandre}, Gloria and {Swartz}, Douglas A. and {Tamagawa}, Toru and {Tavecchio}, Fabrizio and {Taverna}, Roberto and {Tawara}, Yuzuru and {Tennant}, Allyn F. and {Thomas}, Nicholas E. and {Tombesi}, Francesco and {Trois}, Alessio and {Tsygankov}, Sergey S. and {Turolla}, Roberto and {Vink}, Jacco and {Wu}, Kinwah and {IXPE Collaboration}},
        title = "{First Detection of X-Ray Polarization from the Accreting Neutron Star 4U 1820-303}",
      journal = {\apjl},
         year = 2023,
        month = aug,
       volume = {953},
       number = {2},
          eid = {L22},
        pages = {L22},
          doi = {10.3847/2041-8213/acec6e},
archivePrefix = {arXiv},
       eprint = {2306.08476},
 primaryClass = {astro-ph.HE},
       adsurl = {https://ui.adsabs.harvard.edu/abs/2023ApJ...953L..22D}
}

@ARTICLE{Cocchi23,
       author = {{Cocchi}, Massimo and {Gnarini}, Andrea and {Fabiani}, Sergio and {Ursini}, Francesco and {Poutanen}, Juri and {Capitanio}, Fiamma and {Bobrikova}, Anna and {Farinelli}, Ruben and {Paizis}, Adamantia and {Sidoli}, Lara and {Veledina}, Alexandra and {Bianchi}, Stefano and {Di Marco}, Alessandro and {Ingram}, Adam and {Kajava}, Jari J.~E. and {La Monaca}, Fabio and {Matt}, Giorgio and {Malacaria}, Christian and {Miku{\v{s}}incov{\'a}}, Romana and {Rankin}, John and {Zane}, Silvia and {Agudo}, Iv{\'a}n and {Antonelli}, Lucio A. and {Bachetti}, Matteo and {Baldini}, Luca and {Baumgartner}, Wayne H. and {Bellazzini}, Ronaldo and {Bongiorno}, Stephen D. and {Bonino}, Raffaella and {Brez}, Alessandro and {Bucciantini}, Niccol{\`o} and {Castellano}, Simone and {Cavazzuti}, Elisabetta and {Chen}, Chien-Ting and {Ciprini}, Stefano and {Costa}, Enrico and {De Rosa}, Alessandra and {Del Monte}, Ettore and {Di Gesu}, Laura and {Di Lalla}, Niccol{\`o} and {Donnarumma}, Immacolata and {Doroshenko}, Victor and {Dov{\v{c}}iak}, Michal and {Ehlert}, Steven R. and {Enoto}, Teruaki and {Evangelista}, Yuri and {Ferrazzoli}, Riccardo and {Garcia}, Javier A. and {Gunji}, Shuichi and {Hayashida}, Kiyoshi and {Heyl}, Jeremy and {Iwakiri}, Wataru and {Jorstad}, Svetlana G. and {Kaaret}, Philip and {Karas}, Vladimir and {Kislat}, Fabian and {Kitaguchi}, Takao and {Kolodziejczak}, Jeffery J. and {Krawczynski}, Henric and {Latronico}, Luca and {Liodakis}, Ioannis and {Maldera}, Simone and {Manfreda}, Alberto and {Marin}, Fr{\'e}d{\'e}ric and {Marinucci}, Andrea and {Marscher}, Alan P. and {Marshall}, Herman L. and {Massaro}, Francesco and {Mitsuishi}, Ikuyuki and {Mizuno}, Tsunefumi and {Muleri}, Fabio and {Negro}, Michela and {Ng}, Chi-Yung and {O'Dell}, Stephen L. and {Omodei}, Nicola and {Oppedisano}, Chiara and {Papitto}, Alessandro and {Pavlov}, George G. and {Peirson}, Abel L. and {Perri}, Matteo and {Pesce-Rollins}, Melissa and {Petrucci}, Pierre-Olivier and {Pilia}, Maura and {Possenti}, Andrea and {Puccetti}, Simonetta and {Ramsey}, Brian D. and {Ratheesh}, Ajay and {Roberts}, Oliver J. and {Romani}, Roger W. and {Sgr{\`o}}, Carmelo and {Slane}, Patrick and {Soffitta}, Paolo and {Spandre}, Gloria and {Swartz}, Douglas A. and {Tamagawa}, Toru and {Tavecchio}, Fabrizio and {Taverna}, Roberto and {Tawara}, Yuzuru and {Tennant}, Allyn F. and {Thomas}, Nicholas E. and {Tombesi}, Francesco and {Trois}, Alessio and {Tsygankov}, Sergey S. and {Turolla}, Roberto and {Vink}, Jacco and {Weisskopf}, Martin C. and {Wu}, Kinwah and {Xie}, Fei},
        title = "{Discovery of strongly variable X-ray polarization in the neutron star low-mass X-ray binary transient XTE J1701{\ensuremath{-}}462}",
      journal = {\aap},
         year = 2023,
        month = jun,
       volume = {674},
          eid = {L10},
        pages = {L10},
          doi = {10.1051/0004-6361/202346275},
archivePrefix = {arXiv},
       eprint = {2306.10965},
 primaryClass = {astro-ph.HE},
       adsurl = {https://ui.adsabs.harvard.edu/abs/2023A&A...674L..10C}
}

@ARTICLE{Capitanio23,
       author = {{Capitanio}, Fiamma and {Fabiani}, Sergio and {Gnarini}, Andrea and {Ursini}, Francesco and {Ferrigno}, Carlo and {Matt}, Giorgio and {Poutanen}, Juri and {Cocchi}, Massimo and {Mikusincova}, Romana and {Farinelli}, Ruben and {Bianchi}, Stefano and {Kajava}, Jari J.~E. and {Muleri}, Fabio and {Sanchez-Fernandez}, Celia and {Soffitta}, Paolo and {Wu}, Kinwah and {Agudo}, Iv{\'a}n and {Antonelli}, Lucio A. and {Bachetti}, Matteo and {Baldini}, Luca and {Baumgartner}, Wayne H. and {Bellazzini}, Ronaldo and {Bongiorno}, Stephen D. and {Bonino}, Raffaella and {Brez}, Alessandro and {Bucciantini}, Niccol{\`o} and {Castellano}, Simone and {Cavazzuti}, Elisabetta and {Ciprini}, Stefano and {Costa}, Enrico and {De Rosa}, Alessandra and {Del Monte}, Ettore and {Di Gesu}, Laura and {Di Lalla}, Niccol{\`o} and {Di Marco}, Alessandro and {Donnarumma}, Immacolata and {Doroshenko}, Victor and {Dov{\v{c}}iak}, Michal and {Ehlert}, Steven R. and {Enoto}, Teruaki and {Evangelista}, Yuri and {Ferrazzoli}, Riccardo and {Garcia}, Javier A. and {Gunji}, Shuichi and {Hayashida}, Kiyoshi and {Heyl}, Jeremy and {Iwakiri}, Wataru and {Jorstad}, Svetlana G. and {Karas}, Vladimir and {Kitaguchi}, Takao and {Kolodziejczak}, Jeffery J. and {Krawczynski}, Henric and {La Monaca}, Fabio and {Latronico}, Luca and {Liodakis}, Ioannis and {Maldera}, Simone and {Manfreda}, Alberto and {Marin}, Fr{\'e}d{\'e}ric and {Marinucci}, Andrea and {Marscher}, Alan P. and {Marshall}, Herman L. and {Mitsuishi}, Ikuyuki and {Mizuno}, Tsunefumi and {Ng}, C. -Y. and {O'Dell}, Stephen L. and {Omodei}, Nicola and {Oppedisano}, Chiara and {Papitto}, Alessandro and {Pavlov}, George G. and {Peirson}, Abel L. and {Perri}, Matteo and {Pesce-Rollins}, Melissa and {Petrucci}, Pierre-Olivier and {Pilia}, Maura and {Possenti}, Andrea and {Puccetti}, Simonetta and {Ramsey}, Brian D. and {Rankin}, John and {Ratheesh}, Ajay and {Romani}, Roger W. and {Sgr{\`o}}, Carmelo and {Slane}, Patrick and {Spandre}, Gloria and {Tamagawa}, Toru and {Tavecchio}, Fabrizio and {Taverna}, Roberto and {Tawara}, Yuzuru and {Tennant}, Allyn F. and {Thomas}, Nicholas E. and {Tombesi}, Francesco and {Trois}, Alessio and {Tsygankov}, Sergey S. and {Turolla}, Roberto and {Vink}, Jacco and {Weisskopf}, Martin C. and {Xie}, Fei and {Zane}, Silvia},
        title = "{Polarization Properties of the Weakly Magnetized Neutron Star X-Ray Binary GS 1826-238 in the High Soft State}",
      journal = {\apj},
         year = 2023,
        month = feb,
       volume = {943},
       number = {2},
          eid = {129},
        pages = {129},
          doi = {10.3847/1538-4357/acae88},
archivePrefix = {arXiv},
       eprint = {2212.12472},
 primaryClass = {astro-ph.HE},
       adsurl = {https://ui.adsabs.harvard.edu/abs/2023ApJ...943..129C}
}

@ARTICLE{Ursini2023,
       author = {{Ursini}, F. and {Farinelli}, R. and {Gnarini}, A. and {Poutanen}, J. and {Bianchi}, S. and {Capitanio}, F. and {Di Marco}, A. and {Fabiani}, S. and {La Monaca}, F. and {Malacaria}, C. and {Matt}, G. and {Miku{\v{s}}incov{\'a}}, R. and {Cocchi}, M. and {Kaaret}, P. and {Kajava}, J.~J.~E. and {Pilia}, M. and {Zhang}, W. and {Agudo}, I. and {Antonelli}, L.~A. and {Bachetti}, M. and {Baldini}, L. and {Baumgartner}, W.~H. and {Bellazzini}, R. and {Bongiorno}, S.~D. and {Bonino}, R. and {Brez}, A. and {Bucciantini}, N. and {Castellano}, S. and {Cavazzuti}, E. and {Chen}, C. -T. and {Ciprini}, S. and {Costa}, E. and {De Rosa}, A. and {Del Monte}, E. and {Di Gesu}, L. and {Di Lalla}, N. and {Donnarumma}, I. and {Doroshenko}, V. and {Dov{\v{c}}iak}, M. and {Ehlert}, S.~R. and {Enoto}, T. and {Evangelista}, Y. and {Ferrazzoli}, R. and {Garcia}, J.~A. and {Gunji}, S. and {Hayashida}, K. and {Heyl}, J. and {Iwakiri}, W. and {Jorstad}, S.~G. and {Karas}, V. and {Kislat}, F. and {Kitaguchi}, T. and {Kolodziejczak}, J.~J. and {Krawczynski}, H. and {Latronico}, L. and {Liodakis}, I. and {Maldera}, S. and {Manfreda}, A. and {Marin}, F. and {Marinucci}, A. and {Marscher}, A.~P. and {Marshall}, H.~L. and {Massaro}, F. and {Mitsuishi}, I. and {Mizuno}, T. and {Muleri}, F. and {Negro}, M. and {Ng}, C. -Y. and {O'Dell}, S.~L. and {Omodei}, N. and {Oppedisano}, C. and {Papitto}, A. and {Pavlov}, G.~G. and {Peirson}, A.~L. and {Perri}, M. and {Pesce-Rollins}, M. and {Petrucci}, P. -O. and {Pilia}, M. and {Possenti}, A. and {Puccetti}, S. and {Ramsey}, B.~D. and {Rankin}, J. and {Ratheesh}, A. and {Roberts}, O.~J. and {Romani}, R.~W. and {Sgr{\`o}}, C. and {Slane}, P. and {Soffitta}, P. and {Spandre}, G. and {Swartz}, D.~A. and {Tamagawa}, T. and {Tavecchio}, F. and {Taverna}, R. and {Tawara}, Y. and {Tennant}, A.~F. and {Thomas}, N.~E. and {Tombesi}, F. and {Trois}, A. and {Tsygankov}, S.~S. and {Turolla}, R. and {Vink}, J. and {Weisskopf}, M.~C. and {Wu}, K. and {Xie}, F. and {Zane}, S.},
        title = "{X-ray polarimetry and spectroscopy of the neutron star low-mass X-ray binary GX 9+9: An in-depth study with IXPE and NuSTAR}",
      journal = {\aap},
         year = 2023,
        month = aug,
       volume = {676},
          eid = {A20},
        pages = {A20},
          doi = {10.1051/0004-6361/202346541},
archivePrefix = {arXiv},
       eprint = {2306.02740},
 primaryClass = {astro-ph.HE},
       adsurl = {https://ui.adsabs.harvard.edu/abs/2023A&A...676A..20U}
}

@ARTICLE{Farinelli23,
       author = {{Farinelli}, R. and {Fabiani}, S. and {Poutanen}, J. and {Ursini}, F. and {Ferrigno}, C. and {Bianchi}, S. and {Cocchi}, M. and {Capitanio}, F. and {De Rosa}, A. and {Gnarini}, A. and {Kislat}, F. and {Matt}, G. and {Mikusincova}, R. and {Muleri}, F. and {Agudo}, I. and {Antonelli}, L.~A. and {Bachetti}, M. and {Baldini}, L. and {Baumgartner}, W.~H. and {Bellazzini}, R. and {Bongiorno}, S.~D. and {Bonino}, R. and {Brez}, A. and {Bucciantini}, N. and {Castellano}, S. and {Cavazzuti}, E. and {Ciprini}, S. and {Costa}, E. and {Del Monte}, E. and {Di Gesu}, L. and {Di Lalla}, N. and {Di Marco}, A. and {Donnarumma}, I. and {Doroshenko}, V. and {Dov{\v{c}}iak}, M. and {Ehlert}, S.~R. and {Enoto}, T. and {Evangelista}, Y. and {Ferrazzoli}, R. and {Garcia}, J.~A. and {Gunji}, S. and {Hayashida}, K. and {Heyl}, J. and {Iwakiri}, W. and {Jorstad}, S.~G. and {Karas}, V. and {Kitaguchi}, T. and {Kolodziejczak}, J.~J. and {Krawczynski}, H. and {La Monaca}, F. and {Latronico}, L. and {Liodakis}, I. and {Maldera}, S. and {Manfreda}, A. and {Marin}, F. and {Marscher}, A.~P. and {Marshall}, H.~L. and {Mitsuishi}, I. and {Mizuno}, T. and {Ng}, C. -Y. and {O'Dell}, S.~L. and {Omodei}, N. and {Oppedisano}, C. and {Papitto}, A. and {Pavlov}, G.~G. and {Peirson}, A.~L. and {Perri}, M. and {Pesce-Rollins}, M. and {Petrucci}, P.~O. and {Pilia}, M. and {Possenti}, A. and {Puccetti}, S. and {Ramsey}, B.~D. and {Rankin}, J. and {Ratheesh}, A. and {Romani}, R.~W. and {Sgr{\`o}}, C. and {Slane}, P. and {Soffitta}, P. and {Spandre}, G. and {Tamagawa}, T. and {Tavecchio}, F. and {Taverna}, R. and {Tawara}, Y. and {Tennant}, A.~F. and {Thomas}, N.~E. and {Tombesi}, F. and {Trois}, A. and {Tsygankov}, S.~S. and {Turolla}, R. and {Vink}, J. and {Weisskopf}, M.~C. and {Wu}, K. and {Xie}, F. and {Zane}, S.},
        title = "{Accretion geometry of the neutron star low mass X-ray binary Cyg X-2 from X-ray polarization measurements}",
      journal = {\mnras},
         year = 2023,
        month = mar,
       volume = {519},
       number = {3},
        pages = {3681-3690},
          doi = {10.1093/mnras/stac3726},
archivePrefix = {arXiv},
       eprint = {2212.13119},
 primaryClass = {astro-ph.HE},
       adsurl = {https://ui.adsabs.harvard.edu/abs/2023MNRAS.519.3681F}
}

@ARTICLE{1988AdSpR...8b.135S,
       author = {{Shakura}, N.~I. and {Sunyaev}, R.~A.},
        title = "{The theory of an accretion disk/ neutron star boundary layer}",
      journal = {Advances in Space Research},
         year = 1988,
        month = jan,
       volume = {8},
       number = {2-3},
        pages = {135-140},
          doi = {10.1016/0273-1177(88)90396-1},
       adsurl = {https://ui.adsabs.harvard.edu/abs/1988AdSpR...8b.135S}
}

@ARTICLE{1985MNRAS.217..291L,
       author = {{Lapidus}, I.~I. and {Sunyaev}, R.~A.},
        title = "{Angular distribution and polarization of X-ray-burster radiation (during stationary and flash phases)}",
      journal = {\mnras},
         year = 1985,
        month = nov,
       volume = {217},
        pages = {291-303},
          doi = {10.1093/mnras/217.2.291},
       adsurl = {https://ui.adsabs.harvard.edu/abs/1985MNRAS.217..291L}
}

@MISC{heasarc,
       author = {{Nasa Heasarc}},
        title = "{HEAsoft: Unified Release of FTOOLS and XANADU}",
 howpublished = {Astrophysics Source Code Library, record ascl:1408.004},
         year = 2014,
        month = aug,
          eid = {ascl:1408.004},
        pages = {ascl:1408.004},
archivePrefix = {ascl},
       eprint = {1408.004},
       adsurl = {https://ui.adsabs.harvard.edu/abs/2014ascl.soft08004N}
}

@ARTICLE{DiMarco2022,
       author = {{Di Marco}, Alessandro and {Costa}, Enrico and {Muleri}, Fabio and {Soffitta}, Paolo and {Fabiani}, Sergio and {La Monaca}, Fabio and {Rankin}, John and {Xie}, Fei and {Bachetti}, Matteo and {Baldini}, Luca and {Baumgartner}, Wayne and {Bellazzini}, Ronaldo and {Brez}, Alessandro and {Castellano}, Simone and {Del Monte}, Ettore and {Di Lalla}, Niccol{\`o} and {Ferrazzoli}, Riccardo and {Latronico}, Luca and {Maldera}, Simone and {Manfreda}, Alberto and {O'Dell}, Stephen L. and {Perri}, Matteo and {Pesce-Rollins}, Melissa and {Puccetti}, Simonetta and {Ramsey}, Brian D. and {Ratheesh}, Ajay and {Sgr{\`o}}, Carmelo and {Spandre}, Gloria and {Tennant}, Allyn F. and {Tobia}, Antonino and {Trois}, Alessio and {Weisskopf}, Martin C.},
        title = "{A Weighted Analysis to Improve the X-Ray Polarization Sensitivity of the Imaging X-ray Polarimetry Explorer}",
      journal = {\aj},
         year = 2022,
        month = apr,
       volume = {163},
       number = {4},
          eid = {170},
        pages = {170},
          doi = {10.3847/1538-3881/ac51c9},
archivePrefix = {arXiv},
       eprint = {2202.01093},
 primaryClass = {astro-ph.IM},
       adsurl = {https://ui.adsabs.harvard.edu/abs/2022AJ....163..170D}
}
\bibliographystyle{aasjournalv7}



\end{document}